\documentclass[reprint, amsmath,amssymb, aps,superscriptaddress]{revtex4-1}
\usepackage{graphicx}% Include figure files
\usepackage{dcolumn}% Align table columns on decimal point
\usepackage{bm}% bold math
\usepackage[english]{babel} % languages, maybe more than one
\usepackage{lmodern} % use latin modern fonts 
\usepackage[dvipsnames]{xcolor}
\usepackage{hyperref} % turn all your internal references into hyperlinks
\usepackage{natbib} % Journal style bibliography managmenet. Bibliography files in the bibtex syntax are expected
\usepackage{amsmath, amsthm, amscd, amssymb} % the AMS packages:
\usepackage{braket}
\usepackage[T1]{fontenc} % accents and so on
\usepackage{geometry} % for document formatting
\makeindex
\begin{document}

%\preprint{APS/123-QED}

%\title{Laser Noise Heating in Cavity Levitodynamics}% Force line breaks with \\
%\title{Resolved Sideband Cooling in the Presence of Laser Phase Noise}% Force line breaks with \\
\title{Resolved-Sideband Cooling of a Levitated Nanoparticle in the\newline Presence of Laser Phase Noise}
\author{Nadine Meyer} \email{nadine.meyer@icfo.eu}
\affiliation{ICFO Institut de Ciencies Fotoniques, The Barcelona Institute of Science and Technology, 08860 Castelldefels (Barcelona), Spain}

\author{Andr\'{e}s de los Rios Sommer}
\affiliation{ICFO Institut de Ciencies Fotoniques, The Barcelona Institute of Science and Technology, 08860 Castelldefels (Barcelona), Spain}

\author{Pau Mestres}
\affiliation{ICFO Institut de Ciencies Fotoniques, The Barcelona Institute of Science and Technology, 08860 Castelldefels (Barcelona), Spain}

\author{Jan Gieseler}
\affiliation{ICFO Institut de Ciencies Fotoniques, The Barcelona Institute of Science and Technology, 08860 Castelldefels (Barcelona), Spain}

\author{Vijay Jain}
\affiliation{Photonics Laboratory, ETH Z{\"u}rich, 8093 Z{\"u}urich, Switzerland}

\author{Lukas Novotny}
\affiliation{Photonics Laboratory, ETH Z{\"u}rich, 8093 Z{\"u}urich, Switzerland}

\author{Romain Quidant}
\affiliation{ICFO Institut de Ciencies Fotoniques, The Barcelona Institute of Science and Technology, 08860 Castelldefels (Barcelona), Spain}
\affiliation{ICREA-Instituci\'{o} Catalana de Recerca i Estudis Avan\c{c}ats, 08010 Barcelona, Spain}

\date{\today}
%%%%%%%%%%%%%%%%%%%%%%%%%%%%%%%%%%%%%%%%%%%%%%%%%%%%%%%%%%%%%%%%%%%%%%%%%%%%%%%%%%%%%%%%%%%%%%%%%%%%%%%
\begin{abstract}
%\noindent We investigate the limitations of resolved sideband cooling due to phase noise heating in the context of cooling the center-of-mass motion of a levitated nanoparticle in a high-finesse cavity. Despite that phase noise heating is not a fundamental physical limitation, it nevertheless embodies up to today the main limitation in reaching the motional ground state of levitated mesoscopic objects with resolved sideband cooling. We reach minimal COM temperatures comparable to $T_{min}=10$mK at pressure of $P = 3\times 10^{-7}$mbar only limited by phase noise.   
\noindent We investigate the influence of  laser phase noise heating on resolved sideband cooling in the context of cooling the center-of-mass motion of a levitated nanoparticle in a high-finesse cavity. Although phase noise heating is not a fundamental physical constraint, the regime where it becomes the main limitation in Levitodynamics has so far been unexplored and hence embodies from this point forward the main obstacle in reaching the motional ground state of levitated mesoscopic objects with resolved sideband cooling. We reach minimal center-of-mass temperatures comparable to $T_{min}=10$mK at a pressure of $p = 3\times 10^{-7}$mbar, solely limited by phase noise. Finally we present possible strategies towards motional ground state cooling in the presence of phase noise. 
\end{abstract}

%%%%%%%%%%%%%%%%%%%%%%%%%%%%%%%%%%%%%%%%%%%%%%%%%%%%%%%%%%%%%%%%%%%%%%%%%%%%%%%%%%%%%%%%%%%%%%%%%%%%%%%
\pacs{Valid PACS appear here}% PACS, the Physics and Astronomy
\keywords{Optomechanics, Levitodynamics, phase noise heating }%Use showkeys class option if keyword
\maketitle
%%%%%%%%%%%%%%%%%%%%%%%%%%%%%%%%%%%%%%%%%%%%%%%%%%%%%%%%%%%%%%%%%%%%%%%%%%%%%%%%%%%%%%%%%%%%%%%%%%%%%%%
\noindent Among the numerous optomechanical systems, Levitodynamical systems excel with an extreme level of isolation from the environment, rendering Q-factors exceeding $10^8$ \cite{Gieseler2013}. This makes them an attractive alternative to membranes and nanobeams \cite{Tsaturyan2014,Reinhardt2016,Norte2016,Ghadimi2018} for probing macroscopic quantum phenomena at room temperature \cite{Marshall2003,Kleckner2008,Romero-Isart2010,Romero-Isart2011}. In addition, Levitodynamics offers unique possibilities unavailable in conventional clamped systems, including free fall \cite{Hebestreit2018a}, rotation  \cite{Arita2013,Kuhn2017,Monteiro2018,ReneReimann*MichaelDodererErikHebestreitRozennDiehl2018} and engineered potentials \cite{Rondin2017}. These unique features make them ideal candidates for enhanced sensing applications \cite{Hempston2017}, out of equilibrium thermodynamics \cite{Gieseler2018} and matter wave interferometry \cite{Bateman2014,Wan2016}. \\
\noindent Thus far, the motional ground state (GS) of levitated nanoparticles remains elusive. The lowest phonon occupation of tens of phonons, has been achieved with continuous measurement and active feedback cooling \cite{Gieseler2012,Jain2016,Tebbenjohanns2019,Conangla2019,Li2011}. In contrast to these active schemes, passive optomechanical cooling provides a way to cool to the GS without continuous measurement, provided that the cavity linewidth is narrower than the mechanical frequency. This so-called sideband cooling technique was originally developed for atomic systems and in combination with cryongenics it has been used for GS cooling ($n < 1$) in a range of optomechanical systems. \\
First Levitodynamics experiments demonstrated 1D sideband cooling \cite{Kiesel2013,Asenbaum2013,Millen2015} from room temperature down to 0.3K \cite{Fonseca2016}. Here we demonstrate 1D resolved sideband cooling of a levitated nanoparticle reaching temperatures of $T_{min} =10\text{mK}$ at a pressure of $p = 3\times 10^{-7}$mbar, a regime where we will show that phase noise heating is indeed the limiting factor. The phonon occupation of the mechanical oscillator yields $n_{ph} \approx 2100$, an occupation $125\times$ less than in previous experiments \cite{Fonseca2016} and comparable to minimal temperatures reached in coherent scattering \cite{Vuletic2001, Gonzalez-Ballestero2019, Windey2019, Delic2019}. Next to the well-known decoherence due to thermal noise and photon recoil \cite{Jain2016}, we investigate in detail the influence of frequency noise of the cavity field, also called phase noise, on the phonon occupation. Phase noise decoherence has so far been largely overlooked in Levitodynamics \cite{Delic2019} despite being previously observed in other platforms \cite{Schliesser2008,Safavi-Naeini2013} where it seriously complicates the creation of low phonon states \cite{Rabl2009,Diosi2008}. \\
%%%%%%%%%%%%%%%%%%%%%%%%%%%%%%%%%%%%%%%%%%%%%%%%%%%%%%%%%%%%%%%%%%%%%%%%%%%%%%%%%%%%%%%%%%%%%%%%%%%%%%%
% general method
%\noindent   
%In order to reach GS in room temperature experiments with much higher initial phonon numbers ($n_{th}  =  k_B T/(\hbar \Omega_m) \approx 6\times 10^7$), the optomechanical coupling $g$ needs to be greater in comparison to cryogenic systems. Naively this can be achieved by using a stronger driving field but as we will demonstrate experimentally the phase noise heating prohibits lower phonon states as predicted in [Rabl,other guys??]. \\
%\noindent In many cases of laser cooling an internal optical transition is needed which are hard to address or totally absent for more complex atoms, molecules and more massive particles. Nevertheless
\noindent Understanding the limitations of sideband cooling techniques with actively driven cavities is essential for many protocols to generate entanglement \cite{Riedinger2018, Hong2017}, non-classical correlations \cite{Riedinger2016} or achieve coherent quantum control \cite{Verhagen2012}. Controlling the mechanical motion of mesoscopic systems on the single phonon quantum level has been achieved  only recently \cite{Chan2011,Teufel2011}. %This has enabled the generation of quantum squeezed \cite{Purdy2013,Safavi2013} and entangled states \cite{Riedinger2018} of motion. 
\\% and for the application of pulsed optomechanics \cite{Vanner2011}.\\

\begin{figure}
	\begin{center}
		\includegraphics[width=0.49\textwidth]{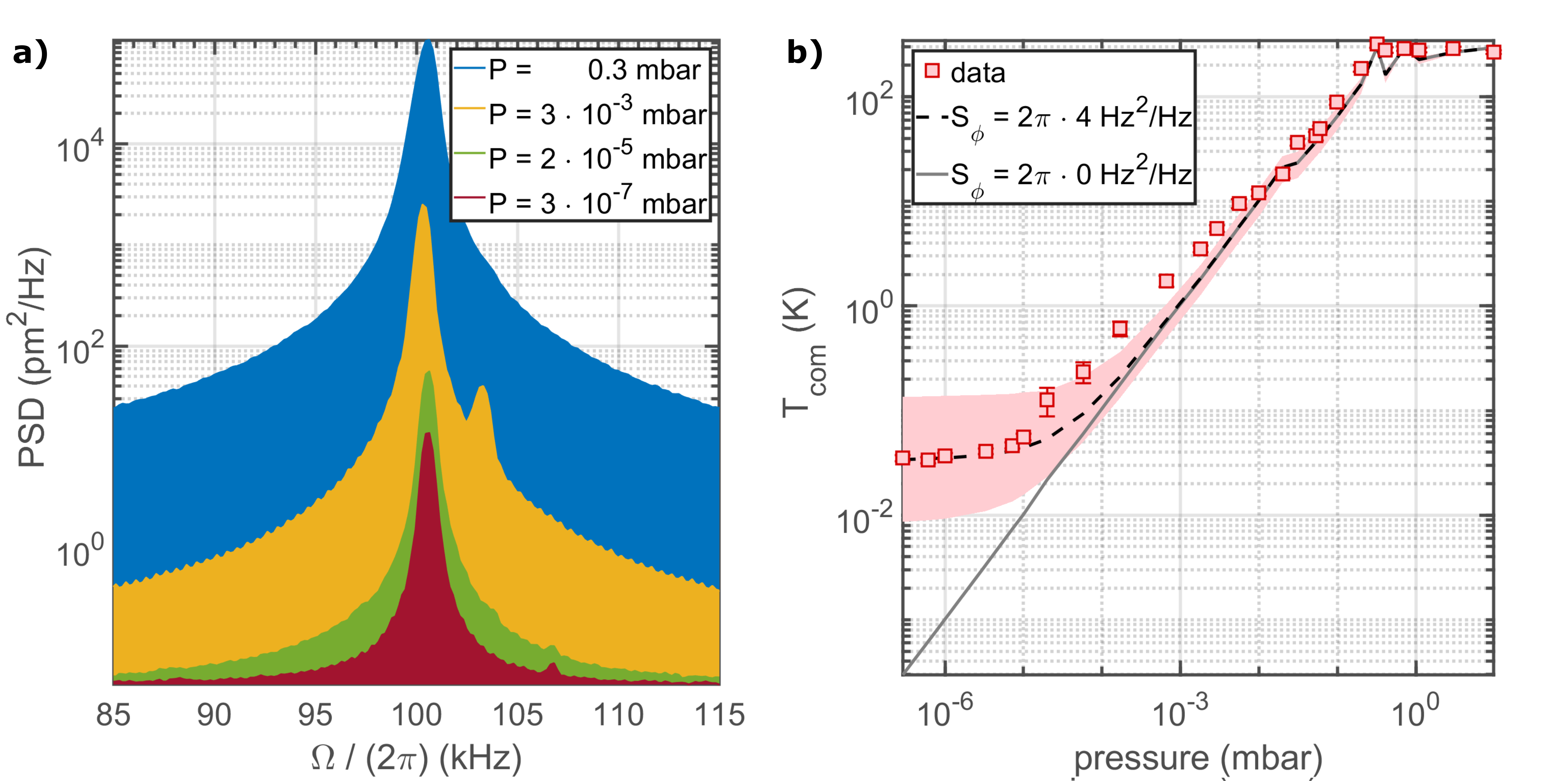}
		\caption{\textbf{Temperature versus Pressure} \textbf{(a)}  PSDs at various pressures ($p = 0.3,\text{ } 3\times 10^{-3}, \text{ }  3\times 10^{-5} \text{ and }  3\times 10^{-7}\text{mbar}$). The area of the PSD $\braket{y^2}$ and therefore the temperature is reduced by reducing the pressure and applying cavity sideband cooling. \textbf{(b)} Applying cavity sideband cooling at optimal detuning $\Delta = -\Omega_y$ and intra cavity power of $P_{intra} = 75$mW. $T_{com}$ reduces linearly with decreasing pressure down to a stable $T_{min} = 35$mK. Theory with negligible phase noise $S_{\phi} = 0 \text{Hz}^2/\text{Hz}$ (solid line) predicts a monotone linear decrease in $T_{com}$ with pressure. Theoretical predictions assuming phase noise of $S_{\phi} =2\pi \times 4\text{Hz}^2/\text{Hz}$ (half-solid line) tails off to a stable final temperature $T_{th} = 34$mK. Shaded area assumes a phase noise regime from half to twice the value of $S_{\phi}$ .}
		\label{fig:2_T_vs_P}
	\end{center}
\end{figure}

\begin{figure}[h]
	\begin{center}
		\includegraphics[width=0.50\textwidth]{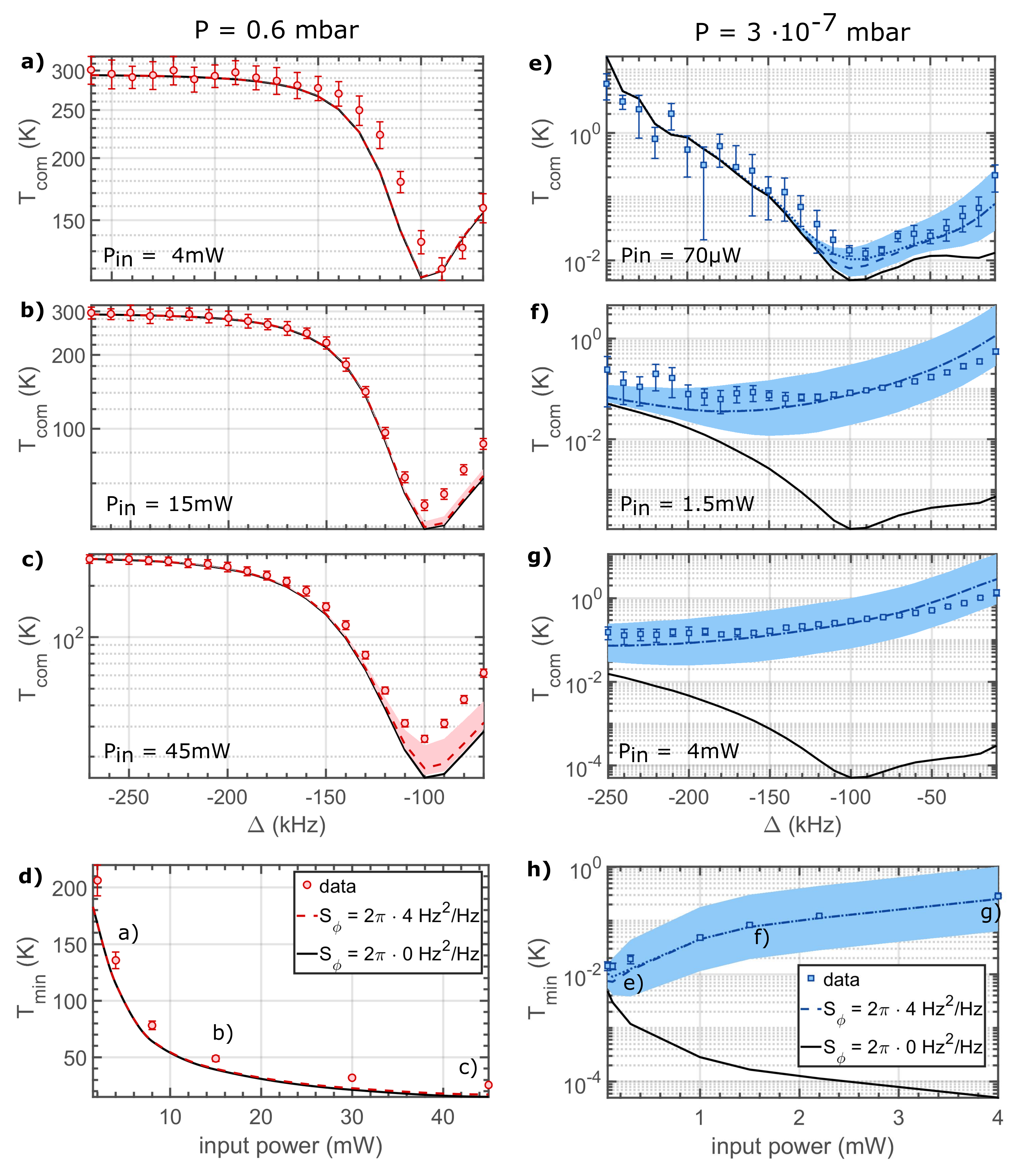}
		\caption{\textbf{Temperature versus detuning $\Delta$} for different input powers $P_{in}$. At high pressures \textbf{(a-d)} clear optimal detuning $\Delta_{opt}\approx -\Omega_m$ is observed. At low pressures  \textbf{(e-h)} the temperature minimum is washed out. Theory assuming $S_{\phi} = 0 \text{Hz}^2/\text{Hz}$ (solid line) predicts stronger cooling with increasing $g$. Theory assuming $S_{\phi} =2\pi \times 4\text{Hz}^2/\text{Hz}$ (half-solid line) accounts for phase noise. The shaded area assumes a phase noise regime from half to twice the value of $S_{\phi}$. In \textbf{(d)} and  \textbf{(h)} $T_{com}$ at optimal detuning versus $P_{in}$ is depicted, showing at low pressure the opposite behaviour in comparison to high pressure.}
		\label{fig:3_T_vs_power}
	\end{center}
\end{figure}

\begin{figure}
	\begin{center}
		\includegraphics[width=0.50\textwidth]{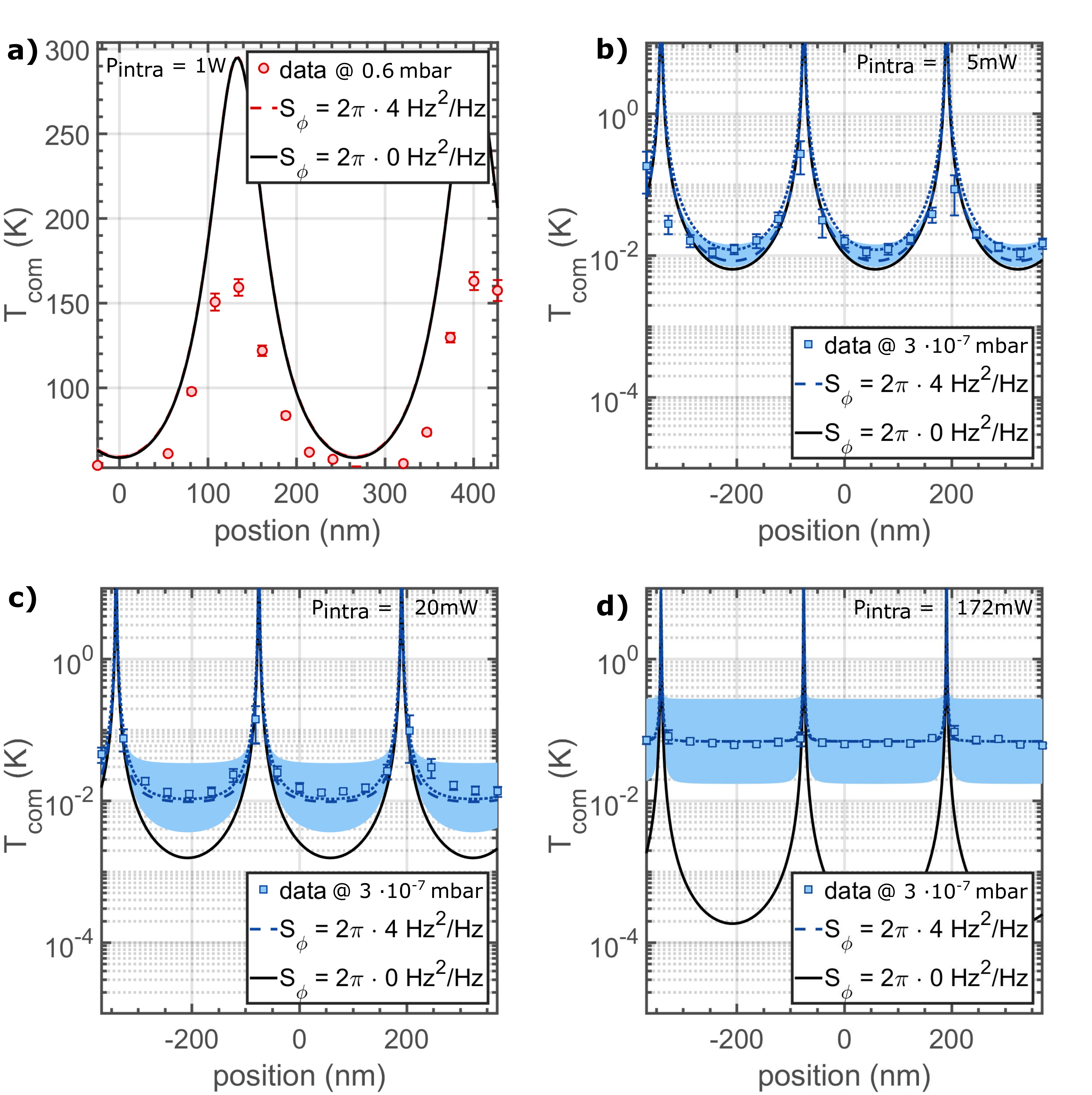}
		\caption{\textbf{Temperature versus particle position $y$} at optimal detuning $\Delta =  - \Omega_m$ and various intracavity powers $P_{intra}$. \textbf{(a)} At high pressures, $T_{com}$ changes sinusoidally with position. The minimal temperature of $T_{min}\approx 50\text{K}$ agrees well with theory (solid line). The maximum temperature deviates from the theory prediction due to the movement of the particle in the cavity field, as discussed in the main text. \textbf{(b} - \textbf{d):} $T_{com}$ at low pressures. For the lowest power \textbf{(b)} $T_{com}$ keeps its sinusoidal dependence on $y$ with a minimal temperature $T_{min} \approx 10\text{mK}$.  The position dependence is gradually lost and the minimum temperature increases, when the intracavity power is raised. Theory neglecting phase noise contributions with $S_{\phi} = 0 \text{Hz}^2/\text{Hz}$ (solid line) consistently predicts a sinusoidal dependence on position $y$. Theory assuming a phase noise level of $S_{\phi} =2\pi \times 4\text{Hz}^2/\text{Hz}$ (half-solid line) accounts for phase noise and for the additional SNRP due to the trap (dotted line). The shaded area assumes a phase noise regime from half to twice the value of $S_{\phi}$ }
		\label{fig:4_T_vs_Pos}
	\end{center}
\end{figure}

\noindent By using an external cavity, the center-of-mass (COM) motion of an atom, ion, molecule \cite{Vuletic2000,Leibrandt2009}, or mesoscopic particle can be controlled and therefore cooled. The presence of a polarizable object inside the cavity induces a position-dependent dispersive change in optical path length%. Therefore the cavity resonance shifts as $\frac{\partial\omega_{cav}}{\partial y}$
, altering the intracavity intensity which then acts back on the particle motion. Coherently driving the cavity with a red(blue) detuned light field enhances(reduces) anti-Stokes scattering versus Stokes scattering, thus cooling(heating) the COM motion.\\
The interaction Hamiltonian for a particle moving along the axis of an optical cavity is $\hat{H}_{int} = -\hbar g_0 \hat{a}^\dagger  \hat{a} (\hat{b} + \hat{b}^\dagger) $ \cite{Aspelmeyer2014,Kiesel2013}
%\begin{equation}\label{eq:Ham}\nonumber
%\hat{H}_{int} = -\hbar g_0 \hat{a}^\dagger  \hat{a} (\hat{b} + \hat{b}^\dagger) 
%\end{equation}
where $\hat{a}$ ($\hat{a}^\dagger$) is the photon annihilation (creation) operator and  $\hat{b}$ ($\hat{b}^\dagger$) is the phonon annihilation (creation) operator. The single photon optomechanical coupling strength $g_0$ can be enhanced by the driving field as $g^2 = g_0^2\hat{a}^\dagger  \hat{a} =  g_0^2n_{cav}$, $n_{cav}$ being the intracavity photon number. The single photon  optomechanical coupling strength is sinusoidally modulated due to the intracavity standing wave and given as %$g_0 = U_0 \sin(2ky) k \sqrt{\frac{\hbar}{2m\Omega_y}}$
\begin{equation}\label{eq:g0}
g_0 = U_0 \sin(2ky) k \sqrt{\frac{\hbar}{2m\Omega_y}}
\end{equation}
where $U_0$ is the resonance frequency shift induced by a particle placed at the center of an empty cavity, with $U_0 = \omega_{cav}\alpha /(2\epsilon_0 V_{cav}) \approx 2\pi\times 10\text{kHz}$, $\omega_{cav}$ being the cavity resonance frequency, $\alpha = 4\pi\epsilon_0 r^3 (n_{p}^2 - 1)/(n_{p}^2 + 2)$ the polarizability of the particle with radius $r = 118\text{nm} \pm 6$nm and refractive index $n_p = 1.45$. The cavity volume is $V_{cav} = \pi  L_{cav} w_{cav}^2/4$, $L_{cav}= 2.43$cm the cavity length, $w_{cav}= 64\mu$m the cavity waist, $k = 2\pi/\lambda_{cav}$ the cavity field wave vector, $\lambda_{cav}= 1064$nm the cavity wavelength  and $y$ the position of the particle from the center along the cavity axis. The particle mass $m = (4/3)\pi r^3 \rho$ is inferred from the particle density $\rho = 2200\text{kg/m}^3$, and the particle mechanical frequency $\Omega_m$ is obtained from the particle displacement power spectral density (PSD). The optomechanical damping rate is then given by \cite{Aspelmeyer2014}
\begin{equation}\label{eq:Gamma_opt}\nonumber
\Gamma_{opt} = g_0^2 n_{cav} \left(\frac{\kappa}{\frac{\kappa^2}{4} + (\Delta + \Omega_y)^2} - \frac{\kappa}{\frac{\kappa^2}{4} + (\Delta - \Omega_y)^2}\right)
\end{equation}
with the cavity linewidth $\kappa= 40$kHz (FWHM). The optomechanical damping rate depends strongly on position along the cavity axis $y$ through $g_0$, the intracavity photon number $n_{cav}$ and detuning from the cavity resonance  $\Delta = \omega_{L} - \omega_{cav}$. In the resolved sideband regime ($\Omega_m \gg \kappa$) the maximum cooling rate equals $\Gamma_{opt} = 4 g_0^2 n_{cav}/\kappa \approx 2\pi\times 2\mu \text{Hz} \: n_{cav}$ at optimal red detuning $\Delta = - \Omega_m$, enabling an optomechanical damping rate in the kHz-regime in state-of-the-art cavities.\\
\noindent In addition to the coupling rate to the thermal bath $\Gamma_m$, shot noise radiation pressure heating (SNRP) due to the cavity field ($\Gamma_{cav}$) and the trapping field ($\Gamma_{t}$) are additional decoherence sources (see Eq.\eqref {eq_Gamma_m} -  \eqref {eq_Gamma_cav}). As shown in section \ref{sec:suppl_phonon}, the additional phonon occupation due to the SNRP of the cavity light field ($n_{rad \:cav} \ll 1$) does not depend on the intra-cavity photon number, while the SNRP of the trapping light field acts as an additional thermal bath. The latter causes only a small relative offset and will therefore be negelected in the following. Moreover, heating effects due to classical laser intensity noise show a much smaller heating effect \cite{VJPrivate} compared to SNRP and will therefore also be neglected. \\
\noindent In the regime where the thermal mechanical damping is the main decoherence source, the final phonon occupation of the mechanical oscillator is  
\begin{equation}\label{eq:n_ph}
n_{ph}= \frac{\Gamma_{opt} n_{min} + \Gamma_{m} n _{th}}{\Gamma_{opt} +\Gamma_{m}} \approx n_{min} + \frac{\Gamma_{m} n _{th}}{\Gamma_{opt}} 
\end{equation}
where $n_{th}  =  k_B T/(\hbar \Omega_m) \approx 6\times 10^7$ is the initial thermal phonon occupation. We neglect the contribution from the thermal photon occupation of the undriven cavity, since  $n_{cav} = \frac{k_B T}{\hbar \omega_{cav}}\ll1 $ for optical frequencies.  $n_{min} $ puts an ultimate limit on the minimum phonon number for $\Gamma_{opt}\gg \Gamma_{m}$. As a consequence the GS can only be reached in the resolved sideband regime ($\Omega_{m} > \kappa$) where $n_{min} = \kappa^2/(4\Omega_m)^2<1$.
The COM temperature is then $T_{com} = n_{ph}\hbar \Omega_m/k_B$ (solid lines in Fig.\ref{fig:2_T_vs_P} - \ref{fig:4_T_vs_Pos}). 
%Note that we understand temperature not as the bulk temperature ($T_{bulk}\geq 295$K).\\ %The single photon cooperativity $C_0$, a measure for the coherence of the system, is given as  $C_0 = g_0^2/(\kappa \Gamma_m) $ while the cavity enhanced cooperativity is given as $C = C_0 n_{cav}$. For performing quantum experiments the cooperativity needs to be $C>1$.\\

\begin{figure}
	\begin{center}
		\includegraphics[width=0.50\textwidth]{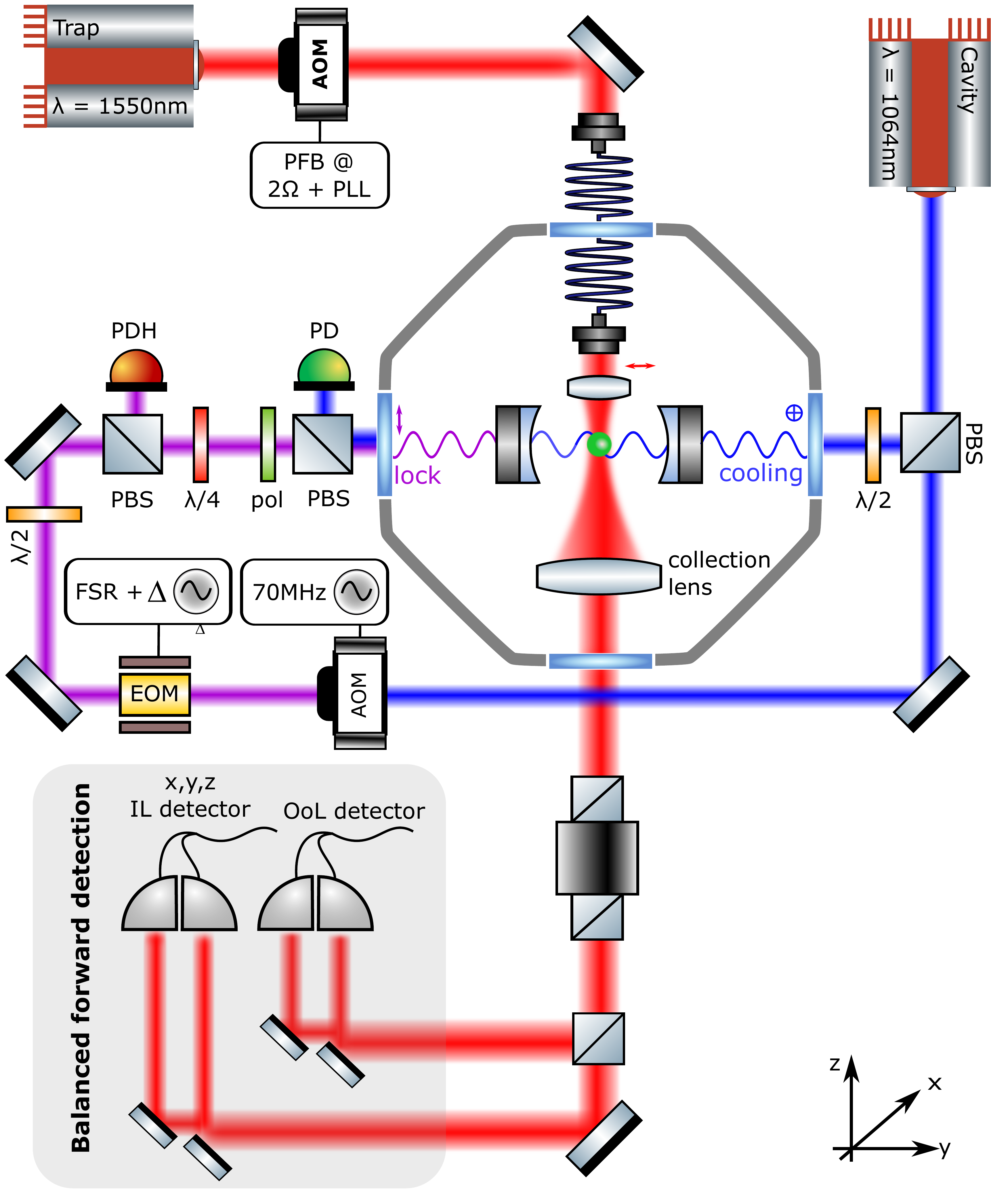}
		\caption{\textbf{Experimental setup} The nanoparticle levitates in a mobile optical tweezers trap (red), positioned in the center of the high finesse cavity field. A weak cavity light field (purple) observed on a photodiode (PDH) is used for Pound-Drever-Hall locking on the cavity resonance $\omega_{cav}$. The cross polarized pump field (blue) is frequency modulated with an EOM at $\text{FSR} + \Delta$ and its transmission is recorded (PD). Standard PFC of the optical tweezer trap prevents particle loss and cross coupling between different degrees of freedom ($x,y,z$). A piezo stage allows for precise 3D positioning of the particle along the cavity axis. The collected trapping light is used in balanced forward detection.}
		\label{fig:1_setup}
	\end{center}
\end{figure}

\noindent In Fig.~\ref{fig:1_setup} the experimental setup is displayed. A silica nanoparticle is levitated in an optical tweezers trap \cite{Mestres2015} with a wavelength $\lambda = 1550~{\rm nm}$, power $P\simeq 185~{\rm mW}$ and focusing lens $\text{NA}=0.8$. The trap is mounted on a 3D piezo system allowing for precise 3D positioning of the particle inside the high finesse Fabry-P\'{e}rot cavity with a cavity finesse $F = 1.55\times 10^5$ and free spectral range FSR = $2\pi\times 6.2$GHz (for more details see supplementary \ref{sec:suppl_exp}). Due to tight focusing, the nanoparticle eigenfrequencies $\Omega_{x,y,z} = 2\pi\times ({\rm 90kHz, 100kHz, 25kHz})$ are non-degenerate. The maximum single photon optomechanical coupling strength is $g_0 = 2\pi\times 0.14\text{Hz}$, which puts GS cooling seemingly into reach by simply increasing the intracavity photon number to $n_{cav} \ge 4.8\times 10^9$, corresponding to a feasible intracavity power of $P_{intra} = 5.5$W.\\ %The single photon cooperativity yields $C_0 = g_0^2/(\kappa \Gamma_m) = 3.5\: 10^{-3}$ at a pressure $p = 3\times 10^{-7}$mbar.\\
\noindent In our experiments we vary the cavity input power $P_{in}$, the detuning  $\Delta$ and the position $y$ along the cavity axis in low and high vacuum, respectively. In the following, points represent data and solid lines are theoretical predictions according to eq.\eqref{eq:n_ph}. The intracavity photon number, used for theoretical predictions,  is calculated from the transmitted cavity power. At low pressure, we apply parametric feedback cooling (PFC) along $x,z$, preventing particle loss and limiting the particle displacement to the linear regime of the optical trap. Experimentally we deduce $T_{com}$ from the area of the particle displacement PSD equal to $\braket{y^2}$ \cite{Hebestreitb}, as shown in Fig.\ref{fig:2_T_vs_P}(a).%, that is proportional to $T_{com}$ . 

\noindent Fig.\ref{fig:2_T_vs_P}(b) shows the pressure dependence of $T_{com}$ at optimal detuning $\Delta = -\Omega_m$ and intracavity power of $P_{intra} = 75\text{mW}$. At pressures below $p < $ 1mbar, we observe the expected linear decrease of $T_{com}$. At  $T_{com} \approx 1\text{K}$, cooling becomes ineffective and the temperature levels off with a constant final minimum temperature of $T_{min}= 35\text{mK}$, in contrast to theoretical expectations (solid line).

\noindent Figs. \ref{fig:3_T_vs_power} and \ref{fig:4_T_vs_Pos} show measurements at high pressure $p = 0.6\text{mbar}$ (red $\circ$) and low pressure $p = 3\times 10^{-7}\text{mbar}$  (blue $\square$), respectively. In Fig.\ref{fig:3_T_vs_power} we investigate $T_{com}$ versus $\Delta$ for various cavity input powers ranging from $P_{in}= 4\text{mW}$ - $45\text{mW}$ at high pressure ($p = 0.6\text{mbar}$) and $P_{in}= 70\mu\text{W}$ - $4\text{mW}$ at low pressure ($p = 3\times 10^{-7}\text{mbar}$). At  high pressure (Fig.\ref{fig:3_T_vs_power}(a-c)) $T_{com}$ features a clear minimum at $\Delta \approx -\Omega_m$. The experimental results agree well with the theory, and only for high cavity input powers of $P_{in} = 45\text{mW}$ we observe a deviation. In contrast, at low pressure the data deviates from the theory and the optimal detuning is farther away from resonance as shown in Fig.\ref{fig:3_T_vs_power}(e-g). Our minimum temperature is $T_{min}\approx 10\text{mK}$, corresponding to a minimal phonon number $n_{min} = 2100$. The dependence of $T_{com}$ at a nominal optimal detuning $\Delta = -\Omega_m$ versus cavity input power is summarized in Fig.\ref{fig:3_T_vs_power}(d) and (h) for high and low pressure respectively. At high pressure $T_{com}$ decreases as expected with increasing power (solid line). This is in strong contrast to the low pressure regime, where measurements deviate from theoretical predictions, which yield a minimal temperature of $T_{th} = 50\mu$K at maximum input power $P_{in} = 4$mW (Fig.\ref{fig:3_T_vs_power}(h)).

\noindent In Fig.\ref{fig:4_T_vs_Pos} we probe $T_{com}$ versus particle position along the cavity axis $y$. We step the optical tweezers trap in increments of $\delta y  =41\text{nm}$ over a total distance exceeding $\lambda_{cav}/2$. The cavity detuning is kept at a constant optimal value of $\Delta = -\Omega_y$ and at constant intracavity power $P_{intra} = 1$W. At high pressure (Fig.\ref{fig:4_T_vs_Pos}(a)) we observe a sinusoidal dependence of the temperature on position as expected from the optomechanical coupling strength $g_0$ (see Eq.\eqref{eq:g0}). While the minimum temperature of $T_{min} = 50K $  agrees well with the theory (solid line), the maximum temperature differs by a factor of 2 from the expected room temperature of $T = 295K$. We attribute this to the particle motion at $T_{com}=160$K, which is $\delta y \approx 20\text{nm}$ and thus a significant fraction of the intracavity standing wave $\lambda_{cav}/2 = 532\text{nm}$.\\
% We attribute this effect to the movement of the particle which at high temperatures is non-negligible in comparison to the spacing of the intracavity standing wave $\lambda_{cav}/2 = 532\text{nm}$. In order to get a reduction of a factor of two in cooling efficiency the particle needs to move by $\delta y = 27\text{nm}$ (shaded grey area) which is comparable to the displacement of $\delta y \approx 20\text{nm}$ corresponding to the maximum $T_{com}$of $T_{max}=160$K. \\
In the low pressure regime the situation is quite different (Fig.\ref{fig:4_T_vs_Pos} (b-d)).  
%Theory predicts again a periodic temperature dependence on position and decreasing temperature with increasing intracavity power (solid lines). 
Periodic behaviour is only observed for the lowest cooling powers $P_{intra}= 5\text{mW}$. Once the intracavity power is increased to $P_{intra}= 20\text{mW}$, $T_{com}$ starts losing its position dependence. The minimum temperature $T_{min} \approx 10$mK persists over a broad region and looses its position-dependence for $P_{intra}= 172\text{mW}$.\\

\noindent Altogether, as long as the dominant heating source is thermal noise, our observations are consistent with theory (see Eq.\eqref{eq:n_ph}). Laser phase noise becomes significant below $p\le 10^{-4}$mbar preventing further cooling. The heating at low pressures cannot be explained by thermal heating (Fig.\ref{fig:2_T_vs_P}) or by photon radiation pressure (see supplementary \ref{sec:suppl_phonon}). \\
\noindent Phase noise stems from a combination of cavity instability and phase noise of the driving laser. It translates into amplitude noise of the intracavity field. This has two effects on the system \cite{Rabl2009}: First, the optomechanical coupling strength $g$ changes due to its dependence on the intracavity photon number $n_{cav}$. However, for a laser linewidth of $\Gamma_L = 1\text{kHz}$ the coupling strength varies as $\kappa\Gamma_L/\Omega_m^2 \ll 1$ and hence the dependence on intracavity field variations is negligible. Second, the conversion of phase to amplitude fluctuations inside the cavity gives rise to a stochastic force driving the mechanical oscillator.
%The additional classical laser intensity noise of the cavity driving field is neglected in the following. 
This leads to an additional phonon occupation $n_{\phi} =  S_{\phi}n_{cav}/\kappa$, which scales linearly with the intracavity photon number $n_{cav}$ and the phase noise PSD at the mechanical frequency $S_{\phi}(\Omega_m)$. %The latter being specific to each experiment. 
Including phase noise, the total final phonon occupation is 
\begin{equation}  \label{eq:n_f}
n_{f}= n_{ph} + n_{\phi} =  \frac{\kappa^2}{16\Omega_m^2} + \frac{\Gamma_{m} k_B T\kappa}{4 g_0^2n_{cav}\;\hbar\Omega_m} +  \frac{S_{\phi}}{\kappa}n_{cav}
\end{equation} 
\noindent where the first two terms derive from Eq.\eqref{eq:n_ph} and the last term accounts for phase noise. Eq.\eqref{eq:n_f} reproduces the data well (half-solid line), assuming the specified phase noise at 10kHz of $S_{\phi} = 2\pi \times 4\text{Hz}^2/\text{Hz}$. The shaded area covers a range of $S_{\phi}/2$ and $2\times S_{\phi}$ to account for the 1/$\Omega$ decrease in phase noise at higher frequencies \cite{Kenji2018} and additional phase noise contributions related to setup instabilities respectively. In general, phase noise heating increases near the cavity resonance due to high intracavity photon numbers (see Eq.\eqref{eq:n_f}) and dominates at low pressure.  This leads to a shift in optimal detuning towards $\Delta < -\Omega_m$ and the opposite power dependence at high and low pressure. The trap SNRP is largely negligible (dotted line in Fig.\ref{fig:3_T_vs_power}(e-h) and Eq.\eqref{eq:n_ph_tot}).
%\begin{equation}  \label{eq:n_f}
%n_{f}= n_{ph} + n_{\phi} =  \frac{\frac{ g_0^2n_{cav}\kappa}{4\Omega_m^2} + \Gamma_{m} \frac{k_B T}{\hbar\Omega_m}}{\frac{4 g_0^2n_{cav}}{\kappa} +\Gamma_{m}} +  \frac{S_{\phi}}{\kappa}n_{cav}
%\end{equation} 
\noindent The optimum intracavity photon number $n_{cav\: opt} = \sqrt{\frac{\kappa^2 \Gamma_m n_{th}}{4 g_0^ 2S_{\phi}}} $
%\begin{equation}\label{eq:opt_ncav}
%n_{cav\: opt} = \frac{\kappa}{4 g_0}\sqrt{\frac{\Gamma_m n_{th}}{S_{\phi}}} + \frac{\Gamma_{m}\kappa}{4g_0^2}
%\end{equation}
 depends on the phase noise level. 
%balancing the cooling (first term of Eq.\eqref{eq:n_f}) and the heating effects (last term of Eq.\eqref{eq:n_f}). 
Consequently, the minimum phonon occupation in presence of phase noise $S_{\phi}$ (see supplementary \ref{sec:suppl_phonon}) is %$n_{ph\: min} = n_{min} + \sqrt{\frac{S_{\phi} \Gamma_m n_{th}}{4 g_0^2}} $.
\begin{equation}\label{eq:opt_nph}
n_{f\: min} = n_{min} + \sqrt{\frac{S_{\phi} \Gamma_m n_{th}}{g_0^2}} 
\end{equation}
%The experimental minimum temperature of $T_{min} = 10$mK stands in good agreement with the theoretical prediction of $T_{f \: min} = 8.4$mK. \\
The experimental minimum phonon occupation of $n_{ph} = 2100$, stands in good agreement with the theoretical prediction of $n_{f \: min} = 1750$, corresponding to $T_{min} = 10$mK and $T_{f\:min} = 8.4$mK respectively. \\
 %The predicted intracavity power is  $P_{intra}=1$0mW which is in reasonable agreement with the experimental region of $P_{intra}=5-20$mW (see Fig.\ref{fig:4_T_vs_Pos}(b-c)).
% If the phonon occupation contributed by the trapping light is neglected, a minimal phonon number of $n_{ph \: min}\approx$1330 can supposedly be reached and the optimal intracavity power decreases. 

%%%%%%%%%%%%%%%%%%%%%%%%%%%%%%%%%%%%%%%%%%%%%%%%%%%%%%%%%%%%%%%%%%%%%%%%%%%%%%%%%%%%%%%%%%%%%%%%%%%%%%%
%\subsection{\label{sec:Conclusions}Conclusions}
\noindent In conclusion, we experimentally and theoretically investigated the influence of phase noise heating in resolved sideband cooling of a levitated nanoparticle in high vacuum where thermal heating is no longer the main limitation. Counter-intuitively, minimum temperatures are achieved at low intracavity power.
Nevertheless, there are two approaches to continue towards GS cooling. Either the optomechanical coupling strength $g$ is increased by using a larger particle, a higher finesse or a smaller cavity volume \cite{Magrini2018}, such that the cooling efficiency per photon improves. Alternatively the coupling to the environment has to be reduced by further lowering the pressure or the system's phase noise (see Eq.\eqref{eq:opt_nph}). Reducing the current phase noise of $\sqrt{S_{\phi}/(2\pi)} = 2\text{Hz}/\sqrt{\text{Hz}}$  by a factor of 1500, GS cooling can be achieved with the experimental parameters given here. This condition can be relaxed by an additional factor of 100 for a larger particle of r = 250nm at a pressure of $P = 10^{-10}$mbar. Note that, phase noise can be decreased with external filtering cavities acting as low pass filters \cite{Jayich2012a,Safavi-Naeini2013}. This reduces the phase noise by several orders of magnitude \cite{Hald2005}, opening up the road to GS cooling with levitated nanoparticles.

\vspace{0.5cm}

\textbf{Acknowledgments}.\hspace{0.2cm} NM, ADLRS, PM, and RQ acknowledge financial support from the European Research Council through grant QnanoMECA (CoG - 64790), Fundaci\'{o} Privada Cellex, CERCA Programme / Generalitat de Catalunya, and the Spanish Ministry of Economy and Competitiveness through the Severo Ochoa Programme for Centres of Excellence in R$\&$D Grants No. SEV-2015-0522 and No. FIS2016-80293-R. This project has received funding from the European Union's Horizon 2020 research and innovation programme under the Marie Sk\l{}odowska-Curie grant agreement No 713729. JG received funding from the European Union's Marie Sk\l{}odowska-Curie program (SEQOO, H2020-MSCA-IF-2014, grant no. 655369) LN and VJ acknowledge support through the NCCR-QSIT program (Grant No. 51NF40-160591) by the Swiss National Science Foundation.

%\include{bibli}
% \bibliographystyle{abbrv}
% \bibliographystyle{unsrt}
% \bibliographystyle{apsrev4-1} 
%\bibliography{ZZ_Publications-LaserNoise,bibVY}

%\include{bibli}
% \bibliographystyle{abbrv}
% \bibliographystyle{unsrt}
\bibliographystyle{apsrev4-1} 
%\bibliography{ZZ_Publications-LaserNoise,bibVY}


%%%%%%%%%%%%%%%%%%%%%%%%%%%%%%%%%%%%%%%%%%%%%%%%%%%%%%%%%%%%%%%%%%%%%%%%%%%%%%%%%%%%%%%

\clearpage
\appendix
\section{Experiment}\label{sec:suppl_exp}% ------------------------------------------------------------------
\noindent The experimental setup is displayed in Fig.~\ref{fig:1_setup}. In order to control the detuning $\Delta = \omega_{cav} - \omega$ between the cavity resonance $\omega_{cav}$ and the driving field $\omega$, we drive the cavity with two laser fields originating from the same laser at $\lambda_{cav} = 1064{\rm nm}$. The first weak cavity field is used for locking the cavity via the Pound Drever Hall technique (PDH) on the TEM01 mode minimising additional heating effects through the photon recoil heating of the cavity lock field. The PDH errorsignal acts on the internal laser piezo and an external AOM. In order to prevent interference effects between light fields we cross polarize the lock and pump field and separate them in frequency space by one free spectral range (FSR) such that the total detuning between lock and pump field yields $\omega_{pump} = \omega_{Laser} - \omega_{AOM} + \text{FSR} - \Delta $. The tunable EOM modulation is provided by a signal generator. The AOM is modulated at a constant frequency. The intracavity power can be deduced from the transmitted cooling light observed on a photo diode (PD) behind the cavity. \\

\noindent For additional control of the other two motional degrees of freedom ($x,z$) we apply standard parametric feedback cooling \cite{Gieseler2012} by modulating the trapping potential with a phase locked loop (PLL) at twice the particle mechanical frequencies $\Omega_{x,z}$ via an AOM.  This maintains the particle motion in the $x$ and $y$ direction in the linear regime, avoiding mechanical cross coupling with the $y$ mode, while keeping the particle trapped at high vacuum. \\

% detection
\noindent All particle information shown is gained in forward detection interfering the scattered light field with the trap reference beam. The highly divergent trap light is recollected with a collection lens ($\text{NA}=0.8$). We use three balanced in-loop detectors (IL) to monitor the oscillation of the particle in all three degrees of freedom and generate the feedback signal for the PFC \cite{Gieseler2012}. Additionally an out-of-loop detector (OoL) in the $y$-direction solely records data, avoiding noise squashing \cite{Conangla2019,Tebbenjohanns2019} and therefore an underestimation of the particle energy.
%vacuum 
The pressure can be varied between atmospheric pressure and  $p \approx 10^{-7}$mbar.

\section{Data acquisition and evaluation }
\noindent The data time traces are acquired at 1MHz acquisition rate with a home-built balanced detector and calibrated at $p=10$mbar. Each data point consists of N = 10 averages of which each one consists again of at least 100 averaged position PSDs of which each is based on individual 2ms time trace, corresponding to a total minimum measurement time of $t=2$sec. In the region of interest of ROI = $\pm$10-20kHz the averaged PSD is summed up $\sum_{-ROI}^{ROI}S_{yy} = <y^2> $ and the corresponding temperature $T_{com} = m \Omega_m^2 <y^2>/k_B$ where $k_B$ is the Boltzmann constant. The error bars are the standard deviation of the N averages. 

\section{Phonon occupation} \label{sec:suppl_phonon}% ------------------------------------------------------------------
\noindent The oscillator mean phonon occupation is governed by its environment. It is coupled to several baths as e.g. the thermal environment, the photon bath of trap and cavity field and the cavity frequency noise leading to decoherence effects. Each of the baths has an individual occupation state $n_i$ and coupling rate $\Gamma_i$ to the resonator. The steady state phonon occupation can be expressed as the following: 
\begin{eqnarray}\label{eq:n_ph_tot}
n_{f}  =& \frac{\sum \Gamma_i n_i}{\Gamma_i} \\ \nonumber
        =& \frac{\Gamma_{opt} n_{min} + \Gamma_m n_{th} + \Gamma_{cav} n_{cav} + \Gamma_{phase} n_{cav} + \Gamma_{t} n_{t}} {\Gamma_{opt} + \Gamma_m + \Gamma_{cav}  + \Gamma_{phase}  + \Gamma_{t} }\\ \nonumber
%       & \approx& \frac{\Gamma_{opt} n_{min} + \Gamma_m n_{th} + \Gamma_{cav} n_{cav} + \Gamma_{phase} n_{cav} + \Gamma_{t} n_{t}} {\Gamma_{opt} } \\ \nonumber
        \approx & \:n_{min}  + \frac{\Gamma_m n_{th}}{\Gamma_{opt}} + \frac{\Gamma_{cav} n_{cav} }{\Gamma_{opt}} +\frac{\Gamma_{phase} n_{cav}}{\Gamma_{opt}} + \frac{\Gamma_{t} n_{t}}{\Gamma_{opt}}\\\nonumber
         = & n_{min}  + n_{m} + n_{rad \: cav } + n_{phase} + n_{rad \: t }
\end{eqnarray}

where the we only consider the case of a cooled oscillator with $(\Gamma_{opt} \gg \Gamma_i)$. In Eq.\eqref{eq:n_ph_tot} the terms from left to right describe the residual phonon occupation due to 1) the quantum back action of the resolved sideband cooling  $n_{min} = \kappa^2 /(4 \Omega_m)^2 \ll 1$,  2) the thermal bath $n_{m}$, 3) the shot noise radiation pressure of the cavity light field $n_{rad \: cav}$, 4) the phase noise of the cavity light field  $n_{phase}$ and 5) the shot noise radiation pressure of the optical tweezers light field $n_{rad \: t}$.\\
In the resolved sideband regime ($\Omega_m \gg \kappa$) we assume a maximum optomechanical coupling rate $\Gamma_{opt} = 4 g_0^2 n_{cav}/\kappa$  at the optimal detuning ($\Delta = - \Omega_m$) \cite{Aspelmeyer2014}. The coupling rate $\Gamma_i$ are given in the following.

\noindent The thermal bath couples as 
\begin{equation}\label{eq_Gamma_m}
\Gamma_m  =  \frac{k_B T }{\hbar Q_m  n_{th}}  = 15.8 \frac{r^2 p}{m v_{gas}}
\end{equation}
where $Q_m= \omega/\Gamma_m$ is the mechanical quality factor, $n_{th} = \frac{k_B T}{\hbar \Omega_{m}}$ the thermal occupation number, $r$ the particle radius, $p$ the gas pressure and $v_{gas} = \sqrt{3 k_B T/m_{gas}}$ the velocity of the residual gas molecules.\\
The phase noise of the driving field is coupled as  
\begin{equation}\label{eq_Gamma_phase}
\Gamma_{phase}  =   \frac{S_{\phi}}{\kappa } \Gamma_{opt}     =  \frac{S_{\phi}}{\kappa^2} 4 g_0^2 n_{cav}\\
\end{equation}
where $S_{\phi}$ is the PSD of the phase noise at the mechanical frequency $\Omega_m$ and $\kappa$ the cavity line width (FWHM).\\
The radiation pressure shot noise from the cavity and trap are coupled to the particle as 
\begin{eqnarray}
\label{eq_Gamma_cav} \Gamma_{cav}  &= \frac{\alpha^2 k_{cav}^5 I_{cav}}{30 \pi e_0^2 c \,m\,  \Omega_m n_{cav} } =& \frac{\alpha^2 k_{cav}^6 c\, x_{zpf}}{120 \pi e_0^2 V_{cav} } \\
\label{eq_Gamma_trap} \Gamma_{t}  &= \frac{\alpha^2 k_{t}^5 I_{t}}{30 \pi e_0^2 c \, m \, \Omega_m n_{t} }  =&  \frac{\alpha^2 k_{t}^5 P_{t}}{15 \pi^2 e_0^2 c\, m \, \Omega_m n_{t} w_{t}^2}  
\end{eqnarray}
From Eq.\eqref{eq:n_ph_tot} to \ref{eq_Gamma_trap} we can see that on one hand $n_{min}\ll 1$ and $n_{rad\: cav}\ll 1$ are independent from the intracavity photon number and will be neglected for the remainder of the manuscript. On the other hand  $n_{rad\: t}$ and $n_m$ decrease and $n_{phase}$ increase linearly with $n_{cav}$. $n_{rad \:t}$ only causes an small offset. Hence there exists an optimal intracavity photon number $n_{cav \: opt}$ where the minimum phonon occupation $n_{f \: min}$ is reached. This stands in contrast to the standard picture of sideband cooling where the phonon occupation monotonically decreases with the number of intracavity photons $n_{cav}$. \\
The optimal intracavity photon number $n_{cav}$ where we reach the lowest phonon occupation is given as 
\begin{equation}\nonumber
\frac{dn_{f}}{dn_{cav}} = \frac{S_{\phi}}{\kappa} - \frac{ \Gamma_m n_{th} + \Gamma_{t} n_{t}}{\Gamma_{opt} \:n_{cav}} = 0
\end{equation}
\begin{eqnarray}
n_{cav \: opt} &=& \sqrt{(\Gamma_m n_{th} + \Gamma_{t} n_{t}) \cdot \frac{\kappa^2}{4 g_0^2 S_{\phi} }} \\
               &=& \sqrt{\frac{X\kappa^2}{4 g_0^2 S_{\phi} }} \approx 7.8 \times 10^6\nonumber %= 7863512
\end{eqnarray}
where $X = \Gamma_m n_{th} + \Gamma_{t} n_{t}$. The thermal environment and optical trap together can be interpreted as a thermal bath with a higher effective 
temperature $T_{\textit{eff}}$ and therefore higher phonon occupation $n_{th\: \textit{eff}}$. The additional phonon contribution due to the trap light only needs to be taken into account at  the lowest $T_{com}$. The lowest phonon occupation is reached when the phonons are equally distributed between the effective thermal bath $T_{\textit{eff}}$ and the phase noise heating ($n_{phase} = n_{rad \: trap} + n_m$). 
\begin{eqnarray}
n_{f \: min} &=& \sqrt{\frac{\Gamma_m^2 n_{th}^2 S_{\phi}}{4 g_0^2 X}}  + \sqrt{\frac{S_{\phi} X}{4 g_0^2}}   + \sqrt{\frac{\Gamma_{t}^2 n_{t}^2 S_{\phi}}{4 g_0^2 X}} \\ \nonumber &\approx& 1750
\end{eqnarray}

\noindent In case of neglecting the trap radiation pressure ($\Gamma_m n_{th} \gg \Gamma_{t} n_{t}$), the minimum reachable phonon occupation deviates by $\approx 15\%$ and yields $n_{f\:min}\approx 1330$.
The theoretically predicted optimal intracavity power is $P_{intra}=1$0mW which is in reasonable agreement with the experimental value of $P_{intra}=5$-$20$mW (see Fig.\ref{fig:4_T_vs_Pos}(b-c)).

\end{document}